\documentclass[prb,superscriptaddress,twocolumn,floatfix]{revtex4-1}
\usepackage{graphicx}
\usepackage[english]{babel}
\usepackage{bm}
\usepackage{amsmath,amsfonts}
\usepackage{amsbsy}
\usepackage[colorlinks=true,linkcolor=blue,urlcolor=blue,citecolor=blue,breaklinks=true]{hyperref}
\usepackage{xcolor}
\usepackage[utf8]{inputenc}
\DeclareUnicodeCharacter{03BC}{\mbox{$\mu$}}
\DeclareUnicodeCharacter{2009}{ }
\usepackage{graphics}
\usepackage{xcolor}
\usepackage{bbm}

\newcommand{\be}{\begin{equation}}
\newcommand{\ee}{\end{equation}}
\def\boldsymbol#1{\mbox{\boldmath$#1$}}
\usepackage[normalem]{ulem}

\newcommand{\kvec}{\mathbf{k}}

\newcommand{\beq}{\begin{equation}}
\newcommand{\eeq}{\end{equation}}
\newcommand{\beqa}{\begin{eqnarray}}
\newcommand{\eeqa}{\end{eqnarray}}
\newcommand{\bea}{\begin{eqnarray}}
\newcommand{\eea}{\end{eqnarray}}
\usepackage{amssymb}
\usepackage{array}
\usepackage{amsmath}
\usepackage{graphicx}\usepackage{graphics}
\usepackage{dcolumn}
\usepackage{bm}\usepackage{varioref}

\usepackage{physics}

\usepackage{float}

\DeclareFontFamily{U}{mathx}{\hyphenchar\font45}
\DeclareFontShape{U}{mathx}{m}{n}{
      <5> <6> <7> <8> <9> <10>
      <10.95> <12> <14.4> <17.28> <20.74> <24.88>
      mathx10
      }{}
\DeclareSymbolFont{mathx}{U}{mathx}{m}{n}
\DeclareFontSubstitution{U}{mathx}{m}{n}
\DeclareMathAccent{\widecheck}{0}{mathx}{"71}
\DeclareMathAccent{\wideparen}{0}{mathx}{"75}

\begin{document}


\title{Higher-order topological superconductivity in a topological metal 1T'-MoTe$_2$}
\author{Sheng-Jie Huang}
\affiliation{Max Planck Institute for the Physics of Complex Systems, N{\"o}thnitzer Str. 38, 01187 Dresden, Germany}
\affiliation{Department of Physics, Condensed Matter Theory Center, and Joint Quantum Institute, University of Maryland, College Park, Maryland 20742, USA}
\author{Kyungwha Park}
\affiliation{Department of Physics, Virginia Tech, Blacksburg, Virginia, 24061 USA}
\author{Yi-Ting Hsu}
\affiliation{Department of Physics, University of Notre Dame, South Bend, IN 46556 USA}
\date{\today}

\begin{abstract}
One key challenge in the field of topological superconductivity (Tsc) has been the rareness of material realization. This is true not only for the first-order Tsc featuring Majorana surface modes, but also for the higher-order Tsc, which host Majorana hinge and corner modes. Here, we propose a four-step strategy that mathematically derives comprehensive guiding principles for the search and design for materials of general higher-order Tsc phases. Specifically, such recipes consist of conditions on the normal state and pairing symmetry that can lead to a given higher-order Tsc state.  
We demonstrate this strategy by obtaining recipes for achieving three-dimensional higher-order Tsc phases protected by the inversion symmetry. Following our recipe, we predict that the observed superconductivity in centrosymmetric MoTe$_2$ is a candidate for higher-order Tsc with corner modes. 
Our proposed strategy enables systematic materials search and design for higher-order Tsc, which can mobilize the experimental efforts and accelerate the material discovery for higher-order Tsc phases. 
\end{abstract}

\maketitle
\textit{Introduction---}
A recent breakthrough in the theory of topological superconductors (Tsc) is that certain crystalline symmetries can protect a rich variety of higher-order Tsc phases \cite{Shiozaki2014,Trifunovic2017,Khalaf2018,Geier2018,Wang2018,ShiozakiIndicator,WTe2HOTsc,Fischer2020,Apoorv2020,Geier2020,Ono2020refined,Zhang2020dirac,SJHuang2021,Ono2021,C2HOTsc,Cheng2022}, featuring different patterns of Majorana corner or hinge modes (see Fig. \ref{fig:majorana}a for an example). These higher-order Tsc phases can intrinsically exist without utilizing proximity effects and are distinct from the well-known first-order Tsc with Majorana edge or surface modes \cite{Read2000,tenfold,Qi2009}. For instance, two-dimensional (2D) superconductors with rotational\cite{Apoorv2020, C2HOTsc} or inversion\cite{Fischer2020,Geier2020, WTe2HOTsc,SJHuang2021} symmetries can host Majorana corner modes, as shown from both mathematical analyses using real-space classification\cite{WTe2HOTsc,C2HOTsc} and numerical observations in a candidate material, monolayer WTe$_2$\cite{WTe2HOTsc}. 

However, despite the rapid theoretical development and extensive experimental efforts made over the past decades, unambiguously confirmed materials hosting either first- or higher-order Tsc phases remain extremely rare, especially beyond 1D. This is partly because, unlike topological insulators, the topology in Tsc phases is determined not only by the electronic band structures but also by the pairing symmetry. Progress in Tsc material discovery is therefore hindered by the lack of systematic guiding principles for material search and design that are derived from fundamental understanding. 

A natural strategy to design the topology of the superconducting state is to identify the necessary normal state properties and pairing symmetry. 
In this regard, there is a sharp contrast between first-order and higher-order Tsc:
To achieve a first-order Tsc, having the correct pairing symmetry alone can be sufficient regardless of whether the normal state is topological, e.g., an order parameter of $\Delta=p_x+ip_y$ is well known to lead to a first-order Tsc with Majorana edge modes \cite{Read2000}. 
In contrast, to achieve a higher-order Tsc, past works have shown that the normal state topology can play an essential role on top of the pairing symmetry \cite{Wang2018,Ahn2020,WTe2HOTsc,Apoorv2020,Zhang2020dirac,SJHuang2021,Jahin2022}. 
{
\begin{figure}[!]
\includegraphics[width=8cm]{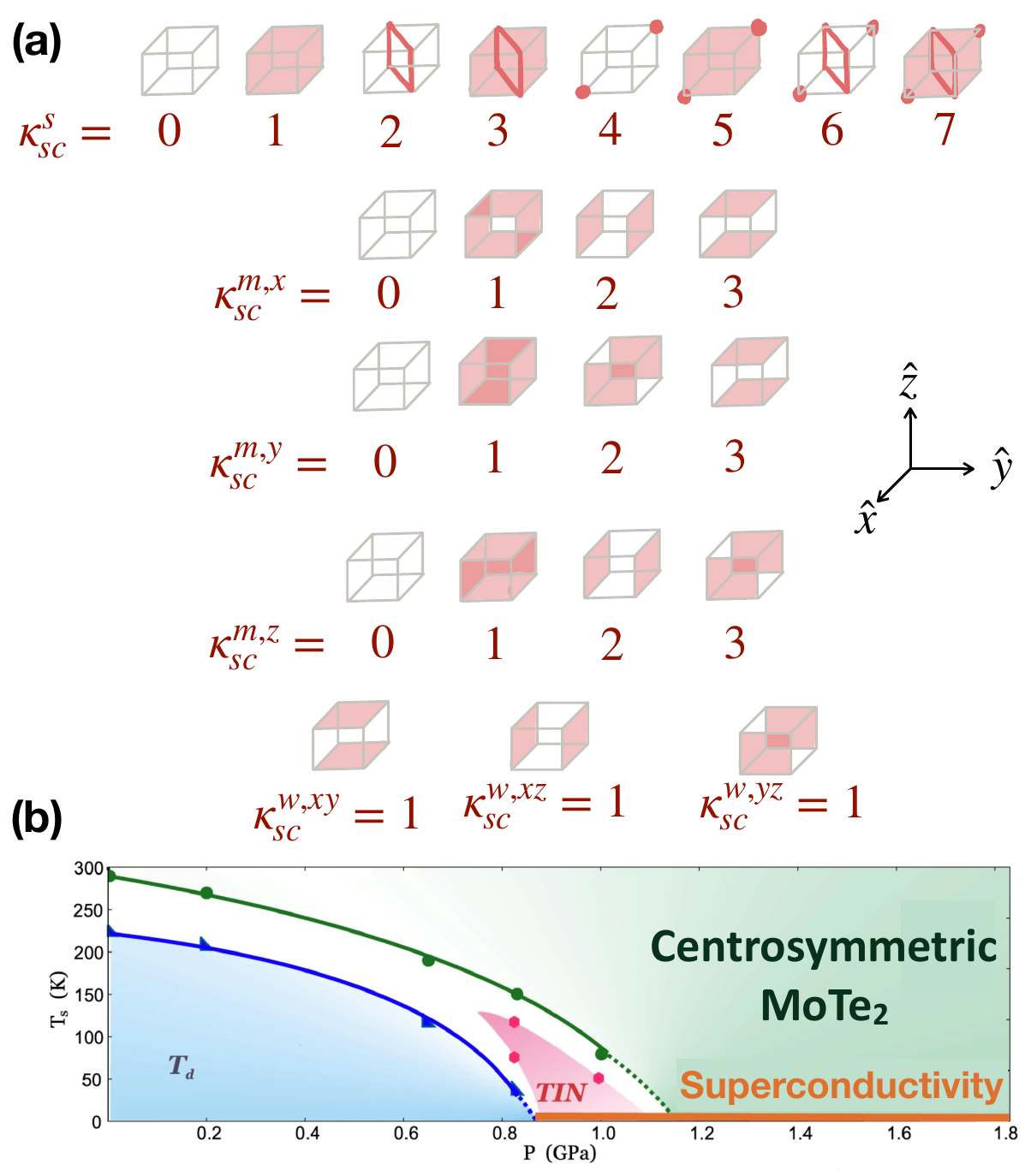}
\caption{(a) Schematics that show the correspondence between our derived symmetry indicators (SIs) $\kappa_{sc}^s$, $\kappa_{sc}^{m,i}$, $\kappa_{sc}^{w,ij}$ and the Majorana boundary patterns (colored by pink) on a cubic geometry for 3D time-reversal superconductors with inversion and translational symmetries. The Majorana patterns and the explicit forms of SIs are obtained in Steps 1 and 2, respectively, and their correspondence is derived from the basis-matching procedure in Step 2. Note that 
the Majorana surface modes associated with the winding number $N_3$ are not shown here. Nonetheless, since the strong SI $\kappa_{sc}^s$ detects the evenness and oddness of $N_3$, the strong phases with odd $\kappa_{sc}^s$ still show one copy of the surface modes. (b) The experimental phase diagram in temperature $T$ and pressure $P$ for MoTe$_2$ adapted from Ref. \onlinecite{Exp_MoTe2}. The experimental finding relevant to this work is the reported superconductivity (shown by the orange line) in the pressure $P$ induced centrosymmetric $T'$ lattice structure (green area). Here, we re-labeled the centrosymmetric lattice structure from ``$T'$'' to ``Centrosymmetric MoTe$_2$" and the superconductivity from ``SC" to ``Superconductivity" for presentation clarity.}
\label{fig:majorana} 
\end{figure}


Here, we propose and demonstrate a four-step strategy to systematically obtain guiding principles for identifying and designing candidate materials of higher-order Tsc. 
Our strategy is described as follows for superconducting materials from a given space group $G$.  
In Step 1, we find all possible first- and higher-order Tsc phases and their Majorana boundary patterns by a real-space classification analysis \cite{Song2017,Huang2018,shiozaki2018generalized,Song2019TC,Shiozaki2019,Song2020,Song2020a} (see Fig. \ref{fig:majorana}a).  
In Step 2, we derive the set of superconducting topological invariants $\kappa_{sc}$ that can diagnose these Majorana boundary patterns from the superconducting band structures, following a protocol developed by some of us\cite{SJHuang2021}. 
In Step 3, we identify the set of topological invariants $\kappa_n$ that characterize the normal-state topology and fermiology. 
In Step 4, we obtain the function $f_{\Delta}$ that relates the two sets of invariants by 
\begin{align}
\kappa_{sc}=f_{\Delta}(\kappa_n)
\label{eq:relation}
\end{align}
for all possible pairing symmetries $\Delta$ in space group $G$. 
From the master equation, Eq. \ref{eq:relation}, guiding principles for the normal state properties described by $\kappa_n$ and pairing symmetry $\Delta$ can be systematically obtained for different higher-order Tsc phases labeled by $\kappa_{sc}$. 

To demonstrate our strategy, an ideal material platform is group-VI transition metal dichalcogenides (TMD), many of which exhibit intrinsic superconductivity with highly tunable normal states exhibiting spin-orbit couplings or band topology\cite{Exp_MoS2sc_Iwasa,Exp_IsingSc_Ye,Exp_NbSe2sc_Mak,Exp_MoS2sc_Pressure,Exp_WTe2_pressure,
Exp_WTe2QSH_Cobden,Exp_WTe2QSH_Pablo,Exp_WTe2sc_Cobden,Exp_WTe2Pressure_Pan,Exp_WTe2sc_Pablo,Li2018}. 
For example, pressure-induced superconductivity was recently discovered in 3D MoTe$_2$ in the centrosymmetric $1T'$ structure\cite{Exp_MoTe2} (see Fig. \ref{fig:majorana}b), where the normal state was proposed to possess higher-order band topology\cite{Tang2019,HOTI_MoTe2}. 
Motivated by the observed superconductivity in 1$T'$-MoTe$_2$, we will demonstrate our strategy on 3D Tsc protected by the inversion symmetry. Following the recipe derived from Eq. \ref{eq:relation}, we predict that 1$T'$-MoTe$_2$ is a plausible higher-order Tsc candidate that hosts corner modes. We support this prediction by a density functional theory (DFT) calculation and microscopic mean-field analysis using a realistic 44-band tight-binding model. 

\textit{Step 1: Majorana boundary patterns---}
Motivated by the superconducting centrosymmetric MoTe$_2$\cite{Exp_MoTe2}, we consider 3D time-reversal superconductors in the simplest inversion-symmetric space group (space group No. 2), which contains the inversion symmetry $\mathcal{I}$ and the three translational symmetries $T_{\hat{r}}$, $r=x,y,z$. Such superconductors are described by a Bogoliubov de Gennes (BdG) Hamiltonian $H_{BdG}=\sum_\kvec\Psi_\kvec^{\dagger}H(\kvec)\Psi_{\kvec}$, where 
\be
H(\kvec)
	 = \begin{pmatrix}
	h(\kvec) & \Delta(\kvec) \\
	\Delta(\kvec)^\dagger & -h(-\kvec)^*
	\end{pmatrix},
	\label{eq:Hbdg} 
\ee
the Nambu basis $\Psi_\kvec=\left( \hat{c}_{\kvec,\uparrow}, \hat{c}_{\kvec,\downarrow}, \hat{c}^\dagger_{-\kvec,\uparrow}, \hat{c}^\dagger_{-\kvec,\downarrow} \right)^\intercal$, and the indices for degrees of freedom other than spin $s=\uparrow,\downarrow$ are suppressed. Here, the normal state $h(\kvec)$ is invariant under the inversion operation $\mathcal{I}(\kvec) h(\kvec) \mathcal{I}(\kvec)^{-1} = h(-\kvec)$ with $\mathcal{I}(-\kvec) \mathcal{I}(\kvec) =1$. For the superconducting gap $\Delta(\kvec)$, we focus on the odd-parity cases where $\mathcal{I}(\kvec) \Delta(\kvec) \mathcal{I}(-\kvec)^{-1} = \eta \Delta(-\kvec)$ with gap parity $\eta=-1$.
Together with the particle-hole and time-reversal symmetries $\mathcal{P}$ and $\mathcal{T}$, the symmetry group of $H(\kvec)$ obeys the following group relations:
\begin{eqnarray}
\mathcal{T}^{2} &=& -1,~\mathcal{P}^{2} =1,~\mathcal{I}_{bdg}^{2} = 1 \nonumber
\\
\left[ \mathcal{T},\mathcal{P} \right] &=& 0,~\left[ \mathcal{T},\mathcal{I}_{bdg} \right] = 0 ,~\{ \mathcal{P}, \mathcal{I}_{bdg} \} = 0,   
\label{eqn:symrelations}
\end{eqnarray}
where $\mathcal{I}_{bdg}=\text{diag}(\mathcal{I},\eta \mathcal{I})$ is the BdG inversion operator that acts on the Nambu basis, and translations simply commute with all other symmetries. Importantly, for odd-parity superconductors, $I_{bdg}$ and $\mathcal{P}$ anticommute so that the particle hole partners have opposite parities. 

To obtain all possible Majorana boundary patterns that a 3D time-reversal centrosymmetric superconductor can support, we first compute the classification group $\mathcal{C}_r$ for crystalline Tsc phases described by $H_{BdG}$, then examine the boundary signature of each phase. 
This can be achieved by using a well-developed real-space classification method called the Topological Crystal Approach\cite{Song2017,Huang2018,shiozaki2018generalized,Song2019TC,Shiozaki2019,Song2020,Song2020a,Zhang2020,SJHuang2021,Zhang2022,One2022realtsc,Zhang2022real3d}. The key idea is that although it is hard to compute $\mathcal{C}_r$ directly in the presence of nonlocal crystalline symmetries, one can dissect the full 3D superconductor into lower-dimensional ``building blocks'' that respect only the local internal symmetries but not the nonlocal crystalline symmetries. Specifically, these building blocks are $d_{b}$-dimensional topological states with $0 \leq d_{b} \leq 3$, where their classification groups and boundary modes have been well studied in the prior literature \cite{tenfold} (see Method section\ref{sec:tc}). By stacking these building blocks into different configurations that respect all the symmetries and checking various consistency conditions \cite{SJHuang2021}, we can determine the Majorana boundary signature of each configuration. These topologically distinct configurations with different Majorana signatures are dubbed \emph{topological crystal states}, where each of them provides a minimal model for each of the Tsc phases. 
Importantly, any 3D superconducting material that respects a given set of crystalline symmetries can be adiabatically connected to a certain topological crystal state \cite{Huang2018,Song2019TC,SJHuang2021}. We therefore expect that the Majorana boundary pattern we obtain for a topological crystal state can also be found in a realistic lattice model for a superconductor in the same Tsc phase. 

The classification group $\mathcal{C}_r$ for our current case of 3D time-reversal superconductors with inversion and translational symmetries is obtained using this approach as follows. First, we identify that the nontrivial building blocks are the time-reversal 1D, 2D, and 3D Tsc states hosting Majorana end, edge, and surface modes, respectively. These building blocks can be stacked on Wycoff positions in different symmetry-allowed configurations to form inversion-symmetric 3D superconducting states with different Majorana boundary patterns. 
By identifying all inequivalent and robust configurations and excluding those that lead to atomic superconductors without Majorana modes, we find that the real-space classification group is given by $\mathcal{C}_r=(\mathbb{Z} \times \mathbb{Z}_{4}) \times (\mathbb{Z}_{4})^{3} \times (\mathbb{Z}_{2})^{3}$ (see Method~section \ref{sec:tc}). 

Next, for each phase captured in $\mathcal{C}_r$, we now discuss the Majorana signature and the protecting symmetries obtained from its building block configuration and will leave the explicit forms of topological invariants to Step 2. 
Specifically, we find that the $\mathbb{Z}$ factor
in $\mathcal{C}_r$ corresponds to first-order strong phases with Majorana surface modes, which can be trivialized by breaking the time-reversal symmetry and is described by a nonzero integer topological invariant $N_3$. 
The first $\mathbb{Z}_{4}$ factor corresponds to inversion-protected higher-order strong phases with Majorana hinge and corner modes. While the $(\mathbb{Z}_{2})^{3}$ corresponds to weak phases protected by two translational symmetries along the $xy$, $yz$, or $xz$-directions, the $(\mathbb{Z}_{4})^{3}$ corresponds to the ``mixed'' phases protected by the inversion and/or translational symmetries. We label them by topological invariants $\kappa_{mixed}^{i}$, $i=x,y,z$, where the $\kappa_{mixed}^{i} = 1$ phases are purely protected by the translational symmetry along $i$-direction, the $\kappa_{mixed}^{i}=2$ phases are protected simultaneously by the inversion and translational symmetries, and the $\kappa_{mixed}^{i} =3$ phases are the stackings of the former two. Note that if we quotient out the phases with an even topological invariant $N_3$, the classification group of the strong phases $\mathbb{Z} \times \mathbb{Z}_{4}$ becomes $\mathbb{Z}_{8}$, consistent with the findings in previous works that did not consider $N_3$ \cite{ShiozakiIndicator,Fischer2020,Ono2020refined}. With this adjustment, the real-space classification group becomes $\mathcal{C}_r = \mathbb{Z}_{8} \times (\mathbb{Z}_{4})^{3} \times (\mathbb{Z}_{2})^{3}$. 

The resulting Majorana signatures for these strong, weak, and mixed Tsc phases are summarized in Fig. \ref{fig:majorana}a, which we obtain by systematically checking the robustness and consistency relations when stacking the building blocks\cite{SJHuang2021} (see the details for this standard procedure of Topological Crystal Approach in Moethod~section \ref{sec:tc}). 
Among these Tsc phases, there are first-order phases with Majorana surface states as well as a rich variety of higher-order Tsc phases with Majorana hinge and corner modes. 

\textit{Step 2: Superconducting state topological invariants $\kappa_{sc}$---} We now turn to the momentum space to derive explicit forms of a set of topological invariants $\kappa_{sc}=\{N_3,\kappa_{sc}^{s},\kappa_{sc}^{m,i},\kappa_{sc}^{w,ij}\}$, $i,j=x,y,z$, which can diagnose the complete Majorana boundary signatures for a given centrosymmetric superconductor (see Fig.~\ref{fig:majorana}a). 
In the following, we show that $N_3$ is the well-known 3D winding number \cite{tenfold,Fu2010,Sato2010,Qi2010} for 3D time-reversal Tsc, while $\kappa_{sc}^{s}$, $\kappa_{sc}^{m,i}$, and $\kappa_{sc}^{w,ij}$ for the strong, mixed, and weak phases are functions of band symmetry data at the high-symmetry points (TRIMs) only. Such invariants are termed symmetry indicators (SIs) \cite{Slager2013,Bradlyn2017,Po2017,Kruthoff2017,Khalaf2018PRX,Watanabe2018,Watanabe2018AZindicators,Ono2019,Fischer2020,Geier2020,Ono2020,Po2020,ShiozakiIndicator,One2020z2}. In particular, we find that the SI for the strong phases $\kappa_{sc}^{s}$ detects both the higher-order phases and the evenness and oddness of $N_3$. 

First, we calculate the momentum space classification group $\mathcal{C}_k$ to determine the nature of each topological invariant. Our calculation is performed using a classification method for topological crystalline phases called Twisted Equivariant K Theory \cite{Shiozaki2022}. We find that the full classification group is given by $\mathcal{K} = \mathbb{Z}^9$, which consists of a subgroup $\mathbb{Z}$ defined on the entire 3D Brillouin zone (BZ) and a subgroup $\mathcal{K'}=\mathbb{Z}^{8}$ restricted only to the TRIMs (see Method~section \ref{sec:kspaceinv}). 
Since this subgroup $\mathbb{Z}$ originates from 3D class-DIII Tsc \cite{tenfold} (see Table III in Method), the phases captured by the $\mathbb{Z}$ subgroup can be labeled by the well-known integer 3D winding number \cite{tenfold,Fu2010,Sato2010,Qi2010} for 3D time-reversal Tsc with Majorana surface modes (see Eq. 36 in Ref. \onlinecite{tenfold}}). On the contrary, the phases in the $\mathcal{K'}$ subgroup are labeled by the SIs $\kappa_{sc}^{s}$, $\kappa_{sc}^{m,i}$, and $\kappa_{sc}^{w,ij}$. 
These phases include strong, weak, and mixed phases with various Majorana signatures, as well as all atomic superconductors, which do not host Majorana boundary modes. 
After removing the contribution from atomic superconductors $\{ \text{AS} \} $, we find that the remaining Tsc phases with non-trivial band topology at TRIMs are classified by $\mathcal{C}_k = \mathcal{K'} / \{ \text{AS} \} = \mathbb{Z}_{8} \times (\mathbb{Z}_{4})^{3} \times (\mathbb{Z}_{2})^{3}$ (see Method section~\ref{sec:kspaceinv}). Therefore, the momentum-space classification $ \mathcal{C}_k$ is consistent with the real-space classification $ \mathcal{C}_r$ we find in Step 1.  

Having shown that the topological invariants that correspond to $\mathcal{C}_k$ are SIs, we now derive the explicit SI expressions that can diagnose the Majorana signatures of the Tsc phases in $\mathcal{C}_k$ (see Fig. \ref{fig:majorana}a). 
Specifically, these SIs were proposed to be different linear combinations of a $\mathbb{Z}$-invariant \cite{ShiozakiIndicator} defined at each of the eight TRIMs $k=\Gamma, X,Y,Z,U,T,R,S$: 
\begin{equation}
\kappa_{sc}^{\eta}=\sum_k\alpha^{\eta}_{k}n_k,~~~ n_{k} = \frac{1}{2} \left( N^{+}[H(k)] - N^{+}[H_{\text{ref}}(k)] \right),
\label{eq:0DinvBdG}
\end{equation}  
where $N^{+}[H^{\text{BdG}}(k)]$ is the number of even-parity occupied states of Hamiltonian $H(k)$, 
and $H_{\text{ref}}(k)$ is a trivial BdG Hamiltonian that serves as a reference point~\cite{SJHuang2021} (see our choice of $H_{\text{ref}}(k)$ in Method section~\ref{Href}). 

The coefficients $\{\alpha^{\eta}_{k}\}$ for these SIs are further obtained by performing a ``basis matching procedure''\cite{SJHuang2021}, where we establish a transparent correspondence between the resulting SIs $\kappa_{sc}^{\eta}$ and the Majorana patterns shown in Fig. \ref{fig:majorana}a. Specifically, to ensure that the $\mathbb{Z}_{8}$, $\mathbb{Z}_{4}$, and $\mathbb{Z}_{2}$ SIs correspond exclusively to the strong, mixed, and weak phases, respectively, we explicitly check the SI values $\kappa_{sc}^{s}$, $\kappa_{sc}^{m,i}$, and $\kappa_{sc}^{w,ij}$ for the real-space minimal models we obtained in Step 1 for each of the strong, mixed, and weak Tsc phases in Fig. \ref{fig:majorana}a (see Method~section  \ref{sec:basis}). These minimal models are models for different topological crystal states built by different building block configurations, where Majorana signatures are evident (see Method~section  \ref{sec:tc}).  
This procedure is necessary because without it, the $\mathbb{Z}_{8}$, $\mathbb{Z}_{4}$, and $\mathbb{Z}_{2}$ SIs in general would each correspond to some profound mixture of strong, mixed, and weak phases due to a basis ambiguity \cite{SJHuang2021}.  
Finally, we arrive at the following SI expressions for time-reversal invariant Tsc phases with inversion and translational symmetries:
\begin{align}
    \kappa_{sc}^{s} &= \sum_{k \in \text{TRIMs}} n_{k} \ \text{mod} \ 8, \nonumber
\\
    \kappa_{sc}^{m,z} &= n_{Z} + n_{U} + n_{T} + n_{R} \ \text{mod} \ 4, \nonumber
\\
    \kappa_{sc}^{m,x} &= n_{X} + n_{S} + n_{U} + n_{R} \ \text{mod} \ 4, \nonumber
\\
    \kappa_{sc}^{m,y} &= n_{Y} + n_{S} + n_{T} + n_{R} \ \text{mod} \ 4, \nonumber
\\
    \kappa_{sc}^{w,xy} &= n_{S} + n_{R} \ \text{mod} \ 2, \nonumber
\\
    \kappa_{sc}^{w,yz} &= n_{U} + n_{R} \ \text{mod} \ 2, \nonumber
\\
    \kappa_{sc}^{w,xz} &= n_{T} + n_{R} \ \text{mod} \ 2, 
\label{eq:indicators}
\end{align}
where the superscripts $s$, $m$, $w$ stand for strong, mixed, and weak phases\footnote{The SIs for this symmetry class have been reported in previos works Ref.~\onlinecite{Ono2020,ShiozakiIndicator} without performing the basis matching procedure.  Therefore, the explicit correspondence between the SIs and the Majorana boundary signatures was not explicitly established in previous works.} 
This set of SIs satisfies the bulk-boundary correspondence so that they can fully distinguish not only all the distinct band topology in the bulk, but also all the Majorana boundary patterns shown in Fig. \ref{fig:majorana}a.
We expect that these SIs are applicable to realistic material-based models since these minimal models are adiabatically connected to any lattice model in the same Tsc phases.   

Before moving on to Step 3, we point out that the complete Majorana signatures are characterized by not just the SIs we find in Eq. \ref{eq:indicators}, but also the 3D winding number $N_3$. In fact, the winding number $N_3$ and SIs $\kappa_{sc}^{\eta}$ are \textit{not} mutually independent. Specifically, the parity of $\kappa_{sc}^{s}$ for strong phases is equal to the winding number $N_3$ modulo $2$\cite{Fu2010,Sato2010,Qi2010}. In Method section~\ref{sec:kspaceinv}, we explicitly show that the pair $(N_{3},\kappa_{sc}^{s})$ is isomorphic to the group $(\mathbb{Z} \times \mathbb{Z}_{4})$, which agrees with the real space classification.

\textit{Step 3: Normal-state invariant $\kappa_n$ ---} 
To characterize the normal state, we adapt the topological invariant for time-reversal topological crystalline insulators (TCI) in the same space group, which contains the inversion $\mathcal{I}$ and translational symmetries $T_{\hat{r}}$. The invariant for such strong TCI phases was proposed to be a $\mathbb{Z}_4$ integer that depends on the electron band parity data at the TRIMs $k$ only\cite{Khalaf2018PRX}:   
\begin{equation}
    \kappa_{I}^s = \frac{1}{4} \sum_{k \in \text{TRIMs}} \left( N^{+}[h_I(k)] - N^{-}[h_I(k)] \right) \ \text{mod} \ 4, 
    \label{eq:kappa0}
\end{equation}
where the superscript $s$ stands for strong phases, $N^{\pm}[h_I(k)]$ is the number of even- and odd-parity occupied bands in the Hamiltonian $h_I(k)$ for the insulator. 
The TCI phases with $\kappa_{strong}^0=1,2,3$ exhibit electronic surface modes, hinge modes, and a combination of both, respectively.      

To adapt this $\mathbb{Z}_4$ invariant $\kappa_{I}^s$ for characterizing the metallic normal states $h(\textbf{k})$ in Eq. \ref{eq:Hbdg}, we now allow it to take both integers and half integers values: 
\begin{equation}
    \kappa_{n}^s = \kappa_{I}^s\vert_{h_{I}(k)\rightarrow h(k)}=0,\frac{1}{2},1,\cdots ,\frac{7}{2}.  
    \label{eq:kappan}
\end{equation} 
Depending on the normal-state fermiology, there are two cases: When all the Fermi surfaces are away from TRIMs, $\kappa^{s}_{n}$ remains a $\mathbb{Z}_4$ integer and the normal state can be viewed as a doped TCI that carries the same band topology as the underlying TCI state.
When the Fermi surfaces circle at least one TRIM, $\kappa^{s}_{n}$ may be a half integer or integer, depending on the number of Fermi pockets circling TRIMs and the topology of the fully occupied bands. For instance, doping a higher-order TCI with hinge modes will lead to a normal state of $\kappa_{n}^s=2$ if Fermi pockets are away from TRIMs. In contrast, a doped trivial insulator with an even-parity electron pocket at $k=\Gamma$ is characterized by $\kappa^{s}_{n}=\frac{1}{2}$\footnote{Note that each band is two-fold degenerate due to the time-reversal and inversion symmetries.}. 
Note that instead of rigorously describing the topology of the normal state, $\kappa_{n}^s$ should be viewed as a computational device for obtaining the recipes in Step 4.  

\textit{Step 4: Recipes for higher-order Tsc states ---}
Equipped with the superconducting and normal state strong invariants $\kappa_{sc}^s$ and $\kappa_n^s$, which are $\mathbb{Z}_8$ and $\mathbb{Z}_4$ numbers respectively, we are now ready to obtain the master equation in Eq. \ref{eq:relation} that relate the two. 
Although $\kappa_{sc}^s$ and $\kappa_{n}^s$ in Eqs. \ref{eq:indicators} and \ref{eq:kappan} are written in terms of the BdG and normal bands, respectively, the relation $f_{\Delta}$ between them can be found in the weak-pairing limit \footnote{The weak-pairing limit is the limit where the gap is smaller than the normal band spacings, which is mostly true for superconductors with low Tc.} where $\kappa_{sc}^s$ can be expressed in terms of the normal band parities for a given pairing symmetry $\Delta$. This is done by expressing the $\mathbb{Z}$-invariant $n_{k}$ at each TRIM $k$ in Eq. \ref{eq:0DinvBdG} in terms of the normal band parities as\cite{ShiozakiIndicator,SJHuang2021} 
\begin{equation}
    n_{k} = \frac{1}{2} \left( N^{+}[h(k)] - N^{-}[h(k)] \right). 
    \label{eq:0DinvNormal}
\end{equation}
Given Eq. \ref{eq:0DinvNormal}, Eq. \ref{eq:indicators}, and Eq. \ref{eq:kappan}, we find that the relating function $f_{\Delta}$ has a simple form 
\begin{equation}
    \kappa_{sc}^s = 2 \kappa^s_{n}, 
\label{eq:strongrelation}
\end{equation}
when the superconducting gap $\Delta$ is parity-odd. In contrast, when the pairing gap $\Delta$ is parity-even, the classification is trivial such that all superconducting invariants $\kappa_{sc}^i$ vanish \cite{Ono2019}. This indicates that an even-parity time-reversal nodeless gap always leads to a topologically trivial superconductor without Majorana modes even when the normal state is topological. 
 

From the relation in Eq.~\ref{eq:strongrelation}, we can deduce recipes for higher-order Tsc phases that consist of conditions on the normal state $\kappa_{n}^s$ in the presence of odd-parity pairing gap $\Delta$. Here, we discuss two example ``recipes''.    
First, if the normal state is a doped strong TI labeled by $\kappa^{s}_{n} = 1$ whose Fermi surfaces are away from TRIMs, introducing an odd-parity gap will drive the system into a second-order Tsc with Majorana hinge modes since $\kappa_{sc}^s=2$. Physically speaking, these Majorana hinge modes are ``leftover'' normal-state surface states that cannot be gapped out by the superconducting gap due to the odd-parity nature.
Second, if the normal state is a doped higher-order TI featuring inversion-protected hinge modes ($\kappa^{s}_{n} = 2$), we expect an exotic third-order Tsc with Majorana corner modes ($\kappa_{sc}^s = 4$) when the system develops an odd-parity pairing gap. 
For the metallic normal state to have $\kappa^{s}_{n} = 2$, the doping-induced Fermi pockets can either be away from TRIMs, or there can be pairs of Fermi pockets that have opposite band parities at TRIMs. 
The latter case is relevant to the MoTe$_2$ case, as shown below. 
For both recipes, the superconducting gap has to be not only odd-parity but also time-reversal symmetric. 
When the Fermi pockets are away from TRIMs, an example gap is a spin-triplet $p_x$-wave gap whose nodal line does not intersect with the Fermi pockets.
When the Fermi pockets circle TRIMs, an example gap is a $^{3}$He-B-phase-like Balian-Werthammer (BW) gap with winding number $N_3=1$: $\Delta_{BW}(\textbf{k})\propto k_x|\uparrow\uparrow-\downarrow\downarrow\rangle+ ik_y|\uparrow\uparrow+\downarrow\downarrow\rangle+ k_z|\uparrow\downarrow+\downarrow\uparrow\rangle$\cite{tenfold,Qi2009}. 
Nonetheless, on top of the Majorana corner modes indicated by $\kappa_{sc}^s = 4$, we also expect Majorana surface modes indicated by the non-zero 3D winding number $N_3$. In a realistic superconductor, depending on the actual hopping parameters and on-site potentials, coexisting Majorana corner and surface modes could experience various levels of hybridization effects.   

\begin{figure}[t]
\begin{center}
\includegraphics[width=0.4 \textwidth]{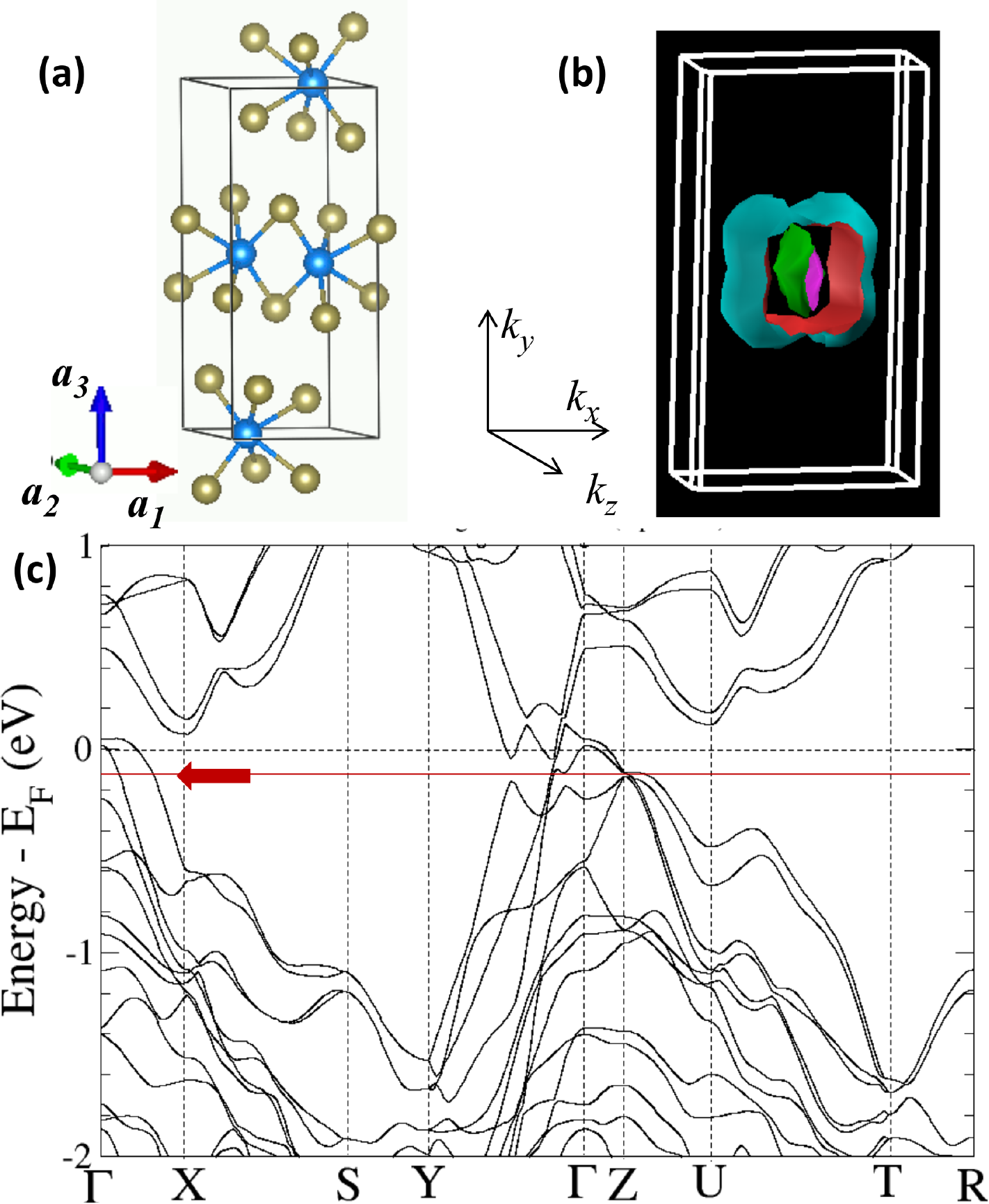}
\caption{(Color online) (a) Crystal structure of MoTe$_2$ (Mo: blue, Te: gray) with the lattice vectors. (b) Fermi surfaces at chemical potential $\mu$$=$$-$46 meV below the Fermi level in the BZ. (c) DFT Band structure without SOC where the $\mu$ value used in (b) is marked with an arrow and a maroon solid line.}
\label{fig:DFT}
\end{center}
\end{figure}

\textit{Symmetry indicators in MoTe$_2$ ---} 
The second recipe suggests that superconducting MoTe$_2$ in the centrosymmetric lattice structure ($1T'$-MoTe$_2$) is a plausible candidate for such a third-order Tsc with Majorana corner modes. 
This is because previous DFT calculations on $1T'$-MoTe$_2$ have reported higher-order band topology along with Fermi pockets located at TRIMs\cite{Tang2019,HOTI_MoTe2}. 
In the following, we will numerically obtain the Majorana boundary signatures in superconducting MoTe$_2$ to examine our prediction made from the last recipe. To this end, we  need to numerically compute the full set of SIs $\{\kappa_{sc}^{s},\kappa_{sc}^{m,i},\kappa_{sc}^{w,ij}\}$, $i,j=x,y,z$ since the existence of mixed and weak phases is also important for determining the full Majorana boudnary signatures. 
Specifically, we perform a DFT calculation on centrosymmetric MoTe$_2$ in an experimentally relevant geometry \cite{Dawson1987} without the spin-orbit coupling, using Vienna
Ab-initio Simulation Package (VASP) \cite{VASP1,VASP2}. A monoclinic primitive unit cell contains 4 Mo atoms and 8 Te atoms (Fig.~\ref{fig:DFT}a). Calculation details are described in Method section~\ref{sec:DFT}. 
We add an on-site Coulomb repulsion (Hubbard $U$) term of 3.0~eV for the Mo $4d$ orbitals, within the 
DFT+U method \cite{Liechtenstein1995}, since it was shown that with this addition, the calculated band structure agree well with
the experimental angle-resolved photoemission spectrum (ARPES) \cite{Exp_MoTe2}. 

When the superconducting gap is small, fully gapped, and parity-odd, which can be energetically favored in the presence of nearest-neighbor attractions, we can obtain the set of SIs $\{\kappa_{sc}^{s},\kappa_{sc}^{m,i},\kappa_{sc}^{w,ij}\}$ following Eq.~\ref{eq:indicators}. 
This is equivalent to obtaining the BdG band parities from the BdG Hamiltonian consisting of a normal state constructed by the DFT bands and a small pairing gap that is fully gapped and parity-odd. 
We find that the full set of SIs is given by 
\begin{align}
\kappa_{sc}^s=4,~~\kappa_{sc}^{m,i}=\kappa_{sc}^{w,ij}=0~~\text{for}~i,j=x,y,z
\label{eq:SI}
\end{align}
at a chemical potential $\mu=-46$~meV below the Fermi level. 
Table~\ref{tab:parity} shows the numbers of bands with positive parity and negative 
parity at the 8 time-reversal invariant momentum (TRIM) points at $\mu$$=$$-$46~meV marked by an arrow in Fig.~\ref{fig:DFT}c. We have checked that the SIs do not change without the $U$ value or with spin-orbit coupling. Besides the computed SIs, there is also a non-zero  winding number $N_3$ given the considered BW pairing gap $\Delta_{BW}$. We expect that $N_3$ is even since the computed strong SI $\kappa_{sc}^s=4$ is even and there are two Fermi pockets at this chemical potential $\mu$ (see Fig.~\ref{fig:DFT}b) that each develops an $N_3=1$ BW gap. 

\begin{table}[!]
\centering
\caption{The numbers of occupied bands $N_e$, positive-parity bands $n_{+}$, and negative-parity bands $n_{-}$ at the 8 TRIM points at $\mu$$=$$-$46~meV for MoTe$_2$. To make the comparison easy, we double the number of bands.}
\begin{tabular}{c|c|c|c|c|c|c|c|c}
\hline\hline
$\mu$$=$$-46$ meV & $\Gamma$ & $X$ & $U$ & $Z$ & $Y$ & $R$ & $S$ & $T$  \\ \hline
      $N_e$ &   68     & 72  &  72 & 68  & 72  & 72  & 72  & 72  \\
      $n_+$ &   38     & 36  &  36 & 34  & 36  & 36  & 36  & 36 \\
      $n_-$ &   30     & 36  &  36 & 34  & 36  & 36  & 36  & 36 \\ \hline \hline
\end{tabular}
\label{tab:parity}
\end{table}
\begin{figure}[!]
\includegraphics[width=8cm]{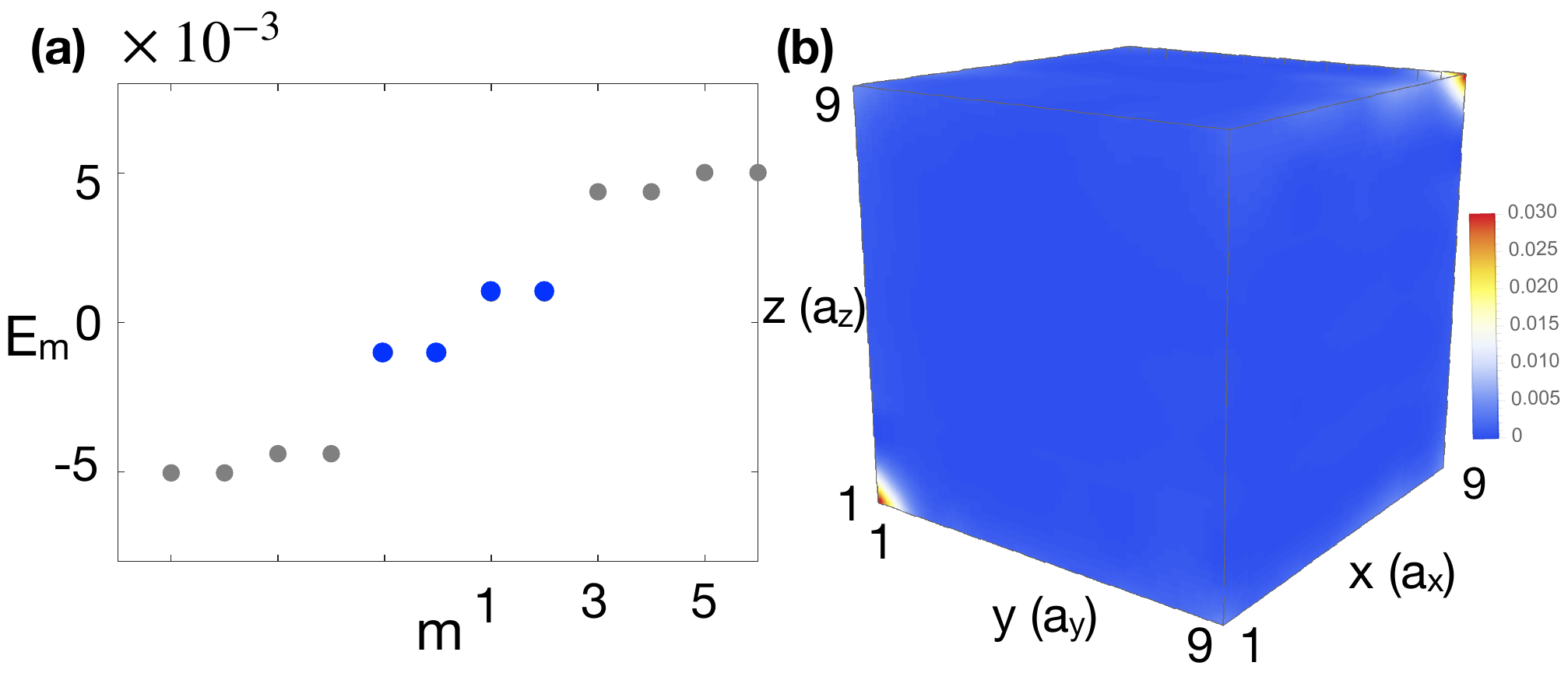}
\caption{(a) The BdG spectrum for MoTe$_2$ at a chemical potential labeled in Fig. \ref{fig:DFT}c and with a BW gap $\Delta_{BW}$ on a finite lattice of $9\times9\times9$ unit cells computed by Krylov method. The gap between the blue states is due to both the finite-size effect and a likely small hybridization between the Majorana corner and surface modes.  (b) The spatial probability distribution $|\psi_m|^2$ of the near-zero-energy BdG eigenstates $\psi_m$ labeled in blue in (a), demonstrating the existence of corner modes when the hybridization is small. The geometry preserves the inversion symmetry, and $a_x$, $a_y$, $a_z$ are the lattice constants in $x$, $y$, $z$ directions, respectively.  
}
\label{fig:corner} 
\end{figure}

\textit{Corner modes in MoTe$_2$---} 
According to the SIs in Eq. \ref{eq:SI},  we expect that centrosymmetric MoTe$_2$ with odd-parity pairing is a higher-order Tsc with Majorana corner modes (see the $\kappa_{sc}^s=4$ phase in Fig. \ref{fig:majorana}a). 
To numerically verify this expectation, we construct a 44-band tight-binding model  for MoTe$_2$ 
based on 44 Wannier functions obtained using WANNIER90 \cite{Marzari1997,Souza2001,Mostofi2008}. The 44 Wannier functions consist of $d_{xy}$, $d_{yz}$, $d_{xz}$, $d_{x^2-y^2}$, and $d_{z^2}$ orbitals 
of all four Mo sites and $p_x$, $p_y$, and $p_z$ orbitals of all eight Te sites in the unit cell.
The constructed tight-binding model reproduces the DFT band structure in the energy range of [-6,3] eV relative to the Fermi level (Fig.~\ref{fig:TB-bands}(a)). Details can be found in Method section~\ref{sec:tb}.

Using this tight-binding model as the normal state, we construct a BdG Hamiltonian with the superconducting gap being the BW gap $\Delta_{BW}$ given in the second recipe in Step 4. We consider the BW gap since the normal state has Fermi surfaces at TRIMs (see Fig. \ref{fig:DFT}b) such that this is the simplest gap structure under which the superconductor is parity-odd but also fully gapped. 
By diagonalizing the constructed BdG Hamiltonian on a 3D open geometry with a system size of $L=9$ \footnote{We are limited to a small system size by the computing power and the not-so-sparse realistic model.}, we find near-zero-energy eigenstates that are localized at a pair of inversion-related corners (see Fig. \ref{fig:corner}), where the choice of which corners is likely determined by the microscopic  positions of Wannier orbitals. 

Importantly, instead of exact zero-energy modes well-separated from other finite-energy quasiparticle states, these low-energy corner modes are \textit{burried in a gapless spectrum} (see Fig. \ref{fig:corner}a). This is expected from the surface modes indicated by the non-zero 3D winding number $N_3$, but not captured by the SIs.  
Consequently, the hybridization between the surface and corner modes can open a small gap and induce some degree of delocalization of the corner modes into the surfaces. This effect is visually not evident in Fig. \ref{fig:corner}b since we choose a chemical potential $\mu$ at which the hybridization strength is likely small to demonstrate the existence of corner modes, but is evident in the gapless spectrum in Fig. \ref{fig:corner}a.  
Therefore, our numerically observed near-zero energy corner modes in Fig. \ref{fig:corner} supports our prediction that 3D centrosymmetric MoTe$_2$ at chemical potential $\mu$ with an odd-parity superconducting gap $\Delta_{BW}$ is a $\kappa_{sc}^s=4$ higher-order Tsc with a non-zero winding number $N_3$. 
We expect that gapped or Majorana corner modes can exist in centrosymmetric MoTe$_2$ with any odd-parity superconducting gap, depending on the coexistence of surface or hinge modes, and such boundary signatures can be detected by Scanning Tunneling Microscope or through transport measurements. 

\textit{Acknowledgement---}Y.-T.H. is grateful for the very helpful discussions with Ruixing Zhang. S.-J.H. acknowledges support from a JQI Postdoctoral Fellowship and the Laboratory of Physical Sciences. Y.-T.H. acknowledges support from NSF Grant No. DMR-2238748. This work was performed in part at Aspen Center for Physics, which is supported by National Science Foundation grant PHY-2210452. This work was supported in part by the National Science Foundation under Grant No. NSF PHY-1748958. 

\appendix
\begin{center}
\textbf{{\large Method}}
\end{center}
\subsection{Topological Crystal Approach.}\label{sec:tc} 
\begin{figure}[!]
\includegraphics[width=8.5cm]{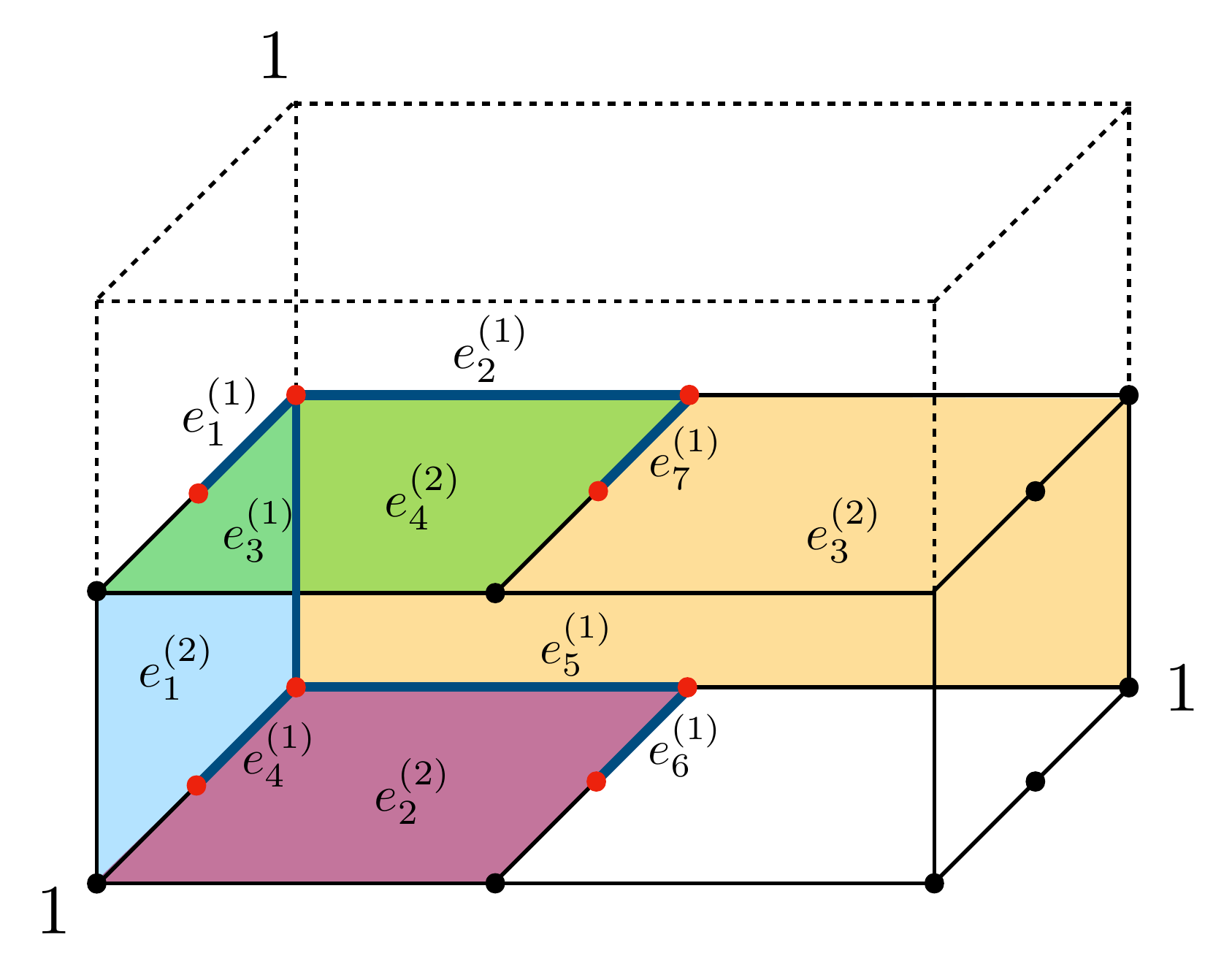}
\caption{Cell-decomposition of the unit cell for space group $P\bar{1}$ ($\# 2$). The origin is chosen at the center of the unit cell. Colored faces are inequivalent 2-cells. Bold blue lines are inequivalent 1-cells. Red dots are inequivalent 0-cells. 
}
\label{fig:realcell} 
\end{figure}

In this section, we explain how we perform the Topological Crystal Approach in Step 1. First, we need to perform a real-space cell-decomposition to break the full unit cell down to 0D, 1D, 2D, and 3D building blocks that do not respect any non-local crystalline symmetries (see Fig.~\ref{fig:realcell}). We denote the dimension of the building block as $d_{b}$. The $d_{b}=3$ building block is the 3d time-reversal invariant topological superconductor in AZ class DIII (3d TSC). Decorating the 3-cells with this building block simply gives the usual 3d TSC with a non-trivial strong indicator $\kappa_{strong}=1$. The $d_{b}=2$ building block is the 2d time-reversal invariant topological superconductor in AZ class DIII (2d TSC). 2d TSCs will be decorated on the 2-cells. The $d_{b}=1$ building block is the 1d time-reversal invariant Kitaev chain (1d TSC), which will be decorated on the 1-cells. Finally, we also have $d_{b}=0$ building block, which is described by a 0d BdG Hamiltonian. The resulting topological crystals and the superconductors that are adiabatically connected to these states are regarded as atomic superconductors (ASC), which are superconducting analogue of atomic insulators. We view such ASC as topologically trivial because they do not host topologically protected boundary zero modes on open geometries.  By quotienting out the ASC, we find that the classification of topological superconductors with non-trivial boundary modes are given by $(\mathbb{Z} \times \mathbb{Z}_{4} )\times \mathbb{Z}_{4}^{3} \times \mathbb{Z}_{2}^{3}$. Table~\ref{tab:tc} shows the decoration patterns of the topological crystals and their corresponding symmetry indicators. The $(\mathbb{Z} \times \mathbb{Z}_{4})$ factor contains the strong first, second and third order TSCs. We denote the strong first, second, and third order TSCs by $(a,b,c)$, where $a$ is an integer corresponding to the strong first order TCS only protected by the internal symmetry in class DIII, $b$ and $c$ are $\mathbb{Z}_{2}$ number corresponding to the strong second and third order TCSs (the first two entries in Table~\ref{tab:tc}). Physically, $a$ is characterized by the 3D winding number. 
They satisfy the following non-trivial stacking relations\cite{Ono2019}:
\begin{align}
(1,0,0) + (1,0,0) &= (2,1,0), \nonumber
\\
(1,0,0) + (-1,0,0) &= (0,1,0), \nonumber
\\
(0,1,0) + (0,1,0) &= (0,0,1).
\label{eq:zxz4}
\end{align}
Note that the phases with even winding numbers are completely decoupled from the higher order phases, i.e. we can freely stack the phases labeled by $(2n,0,0)$ without affecting the higher order phases. One can check that the tuple $(a,b,c)$ satisfying Eq.~\ref{eq:zxz4} is isomorphic to $(\mathbb{Z} \times \mathbb{Z}_{4})$. If we label the group element in $(\mathbb{Z} \times \mathbb{Z}_{4})$ as $(g,h)$, the generator of $\mathbb{Z}$ $(1,0)$ corresponds to the 3d TSC with winding number $1$: $(1,0,0)$, and it's inverse element $(-1,0)$ corresponds the phase $(-1,1,1)$. The $ \mathbb{Z}_{4}$ is generated by the second order phase $(0,1,0)$. Due to the non-trivial stack rules, the phase labeled by $(2,0,0)$ is in fact the $(2,-1)$ element in the group $(\mathbb{Z} \times \mathbb{Z}_{4})$. To better reveal the higher-order topology, it's convenient to quotient out the subgroup generated by $(2,0,0)$ and the resulting group is $\mathfrak{C} = \mathbb{Z}_{8}$ labeled by $(a,b,c)$ but now $a$ is a $\mathbb{Z}_{2}$ number with the stacking rule $(1,0,0) + (1,0,0) \cong (0,1,0)$.  

\begin{table}[htb]
\centering
\begin{tabular}{c|c}
\hline\hline
Symmetry indicators & Decorations 
\\ \hline
         $(2;0,0,0;0,0,0)$ &  $\{ e^{(2)}_{2} + e^{(2)}_{4} \}$    
         \\
         $(4;0,0,0;0,0,0)$ &   $\{ e^{(1)}_{1} + e^{(1)}_{4} + e^{(1)}_{6} + e^{(1)}_{7} \}$
         \\
         \hline
         $(0;1,0,0;0,0,0)$ &   $\{ e^{(2)}_{2}\}$
         \\ 
         $(0;0,1,0;0,0,0)$ &   $\{ e^{(2)}_{1} \}$
         \\
         $(0;0,0,1;0,0,0)$ &   $\{ e^{(2)}_{3}\}$
         \\
         $(0;2,0,0;0,0,0)$ &   $\{ e^{(1)}_{4} + e^{(1)}_{6}\}$
         \\
         $(0;0,2,0;0,0,0)$ &   $\{ e^{(1)}_{1} + e^{(1)}_{4}  \}$
         \\
         $(0;0,0,2;0,0,0)$ &   $\{ e^{(1)}_{2} + e^{(1)}_{5} \}$
         \\
         \hline
         $(0;0,0,0;1,0,0)$ &   $\{ e^{(1)}_{3} \}$
         \\
         $(0;0,0,0;0,1,0)$ &   $\{ e^{(1)}_{4} \}$
         \\
         $(0;0,0,0;0,0,1)$ &   $\{ e^{(1)}_{5} \}$
         \\
         \hline
\end{tabular}
\caption{The topological crystals and symmetry indicators. The first column lists the symmetry indicators for the strong, mixed, and weak phases in the following order: $(\kappa_{strong},\kappa_{mixed}^{z},\kappa_{mixed}^{x},\kappa_{mixed}^{y},\kappa_{weak}^{xy},\kappa_{weak}^{yz},\kappa_{weak}^{zx})$. The second list the topological crystals, presenting as the decoration of the building blocks on the $p$-cells.}
\label{tab:tc}
\end{table}

\subsection{Momentum space topological invariants.}\label{sec:kspaceinv} 
In this section, we discuss our calculation and results in Step 2, where we obtain the momentum-space topological invariants by calculating the equivariant K group $^\phi K_G^{(\tau,c),-3}(BZ)$ using the Atiyah-Hirzebruch Spectral Sequence (AHSS)\cite{Shiozaki2022,SJHuang2021,C2HOTsc}. Please see Ref.~\onlinecite{Shiozaki2022,SJHuang2021,C2HOTsc} for an introduction to the well-developed Equivariant K Theory and AHSS. Here, we present the essential results of our calculation. The elements $E_{2}^{p,-n}$ in the $E_{2}$ page of the AHSS are summarized in Table~\ref{tab:E2}. 

\begin{table}[h]
\begin{center}
\begin{tabular}{c c | c | c | c | c}
	\hline
	AZ class & $n$ & $p=0$ & $p=1$ & $p=2$ & $p=3$
	\\ \hline
	DIII & 3 & $\mathbb{Z}^{8}$ & $1$ & & 
	\\ \hline
	AII & 4 & $\mathbb{Z}^{9}$ & $1$ & &  
	\\ \hline
	CII & 5 &  &  & $1$ & 
	\\ \hline
	C & 6 & & & & $\mathbb{Z}$
	\\ \hline
\end{tabular}
\end{center}
\caption{The $E_2$ page we find from our calculation. According to the Topological Phenomena Interpretation \cite{SJHuang2021}, the three diagonal entries represent the K groups restricted on $p = $ 0-, 1-, 2, and 3-cells for the 3D class-DIII superconductors with inversion symmetry. These three entries together give rise to the full K group.}
\label{tab:E2}
\end{table}

To show that the $E_{2}$ page is the limiting page $E_{\infty}$, one has to calculate the third differential, which is a non-trivial task. We give a physical argument below to show that the $E_{2}$ page is the limiting page. 

Each element in the $E_{2}$ page can be characterized by topological invariants defined in the subspaces of BZ. Topological invariants on 0-cells are classified by $E_{2}^{0,-3} = \mathbb{Z}^{8}$. The explicit form of each $\mathbb{Z}$ invariant is defined in Eq.~\ref{eq:0DinvBdG}. By quotienting out the contribution of ASC, they give the symmetry indicators. There is a $\mathbb{Z}$ invariant defined on the 3-cell. This can be naturally identified with the 3D winding number $N_{3}$\cite{tenfold} that characterized the 3D TSC in class DIII. It has been shown that the parity of the strong symmetry indicator $\kappa_{sc}^{s}$ agrees with the 3D winding number $N_{3}$ modulo $2$\cite{Fu2010,Sato2010,Qi2010}:
\begin{equation}
\kappa_{sc}^{s} \ \text{mod} \ 2 = N_{3} \ \text{mod} \ 2 
\label{eq:N3relation}
\end{equation}
This is consistent with the conjecture that the $E_{2}$ page is the limiting page. Moreover, since there is a non-trivial relation between $\kappa_{sc}^{s}$ and $N_{3}$, the formation group extension that we use to obtain the full $\mathcal{K}$ group has to be non-trivial:
\begin{equation}
1 \rightarrow \mathbb{Z} \rightarrow \mathcal{K} \rightarrow \mathbb{Z}^{8}  \rightarrow 1.
\end{equation} 
Upon quotienting out the contribution of ASC, the $\mathcal{K}$ group should agree with the classification from the Topological Crystal Approach: $(\mathbb{Z} \times \mathbb{Z}_{4} )\times \mathbb{Z}_{4}^{3} \times \mathbb{Z}_{2}^{3}$ in real space. The mixed SIs $\kappa_{sc}^{m,x}$, $\kappa_{sc}^{m,y}$, $\kappa_{sc}^{m,z}$ and weak indicators $\kappa_{sc}^{w,xy}$, $\kappa_{sc}^{w,yz}$, $\kappa_{sc}^{w,xz}$ simply correspond to the $\mathbb{Z}_{4}^{3} \times \mathbb{Z}_{2}^{3}$ factor. The $(\mathbb{Z} \times \mathbb{Z}_{4} )$ factor that corresponds to the strong phases requires further discussion. The strong SI $\kappa_{sc}^{s}$ itself is a $\mathbb{Z}_{8}$ number. However, strong phases are, in fact, characterized by a pair of invariants $(N_{3},\kappa_{sc}^{s})$ with the constraint Eq.~\ref{eq:N3relation}. Taking into account the constraint in Eq.~\ref{eq:N3relation}, the pair $(N_{3},\kappa_{sc}^{s})$ can be parametrized as $(2n+(\kappa_{sc}^{s} \ \text{mod} \ 2), \kappa_{sc}^{s})$, where $n$ is an integer. One can check that the pair $(2n+(\kappa_{sc}^{s} \ \text{mod} \ 2), \kappa_{sc}^{s})$ indeed satisfies the group multiplication rule of $(\mathbb{Z} \times \mathbb{Z}_{4} )$. If we quotient out the subgroup generated by an even winding number $(N_{3},\kappa_{sc}^{s})=(2n,0)$, we obtain $((\kappa_{sc}^{s} \ \text{mod} \ 2), \kappa_{sc}^{s}) \in \mathbb{Z}_{8}$. 

\subsection{Reference Hamiltonian $H_{\text{ref}}$}\label{Href}
In this section, we review why a reference Hamiltonian $H_{\text{ref}}$ is needed\cite{Fischer2020,SJHuang2021} and how we make the choice of $H_{\text{ref}}$. 
In K theory, a K group can be viewed as a formal difference between two vector bundles $B_{1}$ and $B_{2}$. In Karoubi's formulation, the difference is represented by a triple $[B, H_1, H_2]$, where $B$ is the vector bundle whose base space is the BZ and vector space is formed by the occupied states of the Hamiltonian $H_i$, and $H_{1}$ and $H_{2}$ are the flattened Hamiltonians. 

We can further associate different equivalence classes of triples $[B, H, H_{\text{ref}}]$ with distinct gapped phases of matter. 
To do so, instead of using different reference Hamiltonians $H_2$ for different triples, it is crucial to define a `trivial' Hamiltonian $H_{\text{ref}}$ as the universal reference Hamiltonian (i.e., set $H_2=H_{\text{ref}}$) for all triples. For superconductors, a natural choice for the reference Hamiltonian is BdG Hamiltonian formed by a vacuum state, 
\begin{align}
  H_{\text{ref}} = \rm{diag}(\mathbb{I}_N, -\mathbb{I}_N),  
  \label{eq:H0}
\end{align}
, where $N$ is the number of normal bands. In Eq.~\ref{eq:0DinvBdG}, we always use Eq.~\ref{eq:H0} as the reference Hamiltonian. 

\subsection{Basis ambiguity of symmetry indicators}
\label{sec:basis}
In this section, we discuss the basis ambiguity in the calculation of SIs and how to determine the basis for SIs such that the SIs have a transparent correspondence to the Majorana boundary signatures.  SIs are defined as elements in the quotient group $ X = \mathcal{K}' / \{ \text{AS} \}$, where $\mathcal{K}' = \mathbb{Z}^8$ is the classification of the topological invariants defined at TRIMs (the winding number is not included here since the atomic superconductors have a zero winding number), and $\{ \text{AS} \} = \mathbb{Z} \times 8\mathbb{Z} \times (4\mathbb{Z})^{3} \times (2\mathbb{Z})^{3}$ is the classification group of the topological invariants for the atomic superconductors. More specifically, it can be written as a matrix:
\begin{equation}
    M_{\text{AS}} = \begin{pmatrix}
    \boldsymbol{a}_{1} & \boldsymbol{a}_{2} & \boldsymbol{a}_{3} & \boldsymbol{a}_{4} & \boldsymbol{a}_{5} & \boldsymbol{a}_{6} & \boldsymbol{a}_{7} & \boldsymbol{a}_{8}
    \end{pmatrix},
\end{equation}
where each column vector contains the set of 0d invariants $n_{k}$ at TRIMs generated by a atomic superconductor sitting at a Wyckoff position in the real space. The explicit matrix form can be found in Ref.~\cite{ShiozakiIndicator}. To proceed, we compute the Smith normal form to find the linearly independent bases:
\begin{equation}
UM_{\text{AS}}V= \lambda 
\label{eq:snf}
\end{equation}
, where $U$ and $V$ are the transformation matrices for the momentum-space and real-space bases respectively, and $\lambda$ is a diagonal matrix:
\begin{align}
    \lambda  =  \begin{pmatrix}
    1 & & &  
    \\
      & 2I_{3} & & 
    \\
      & & 2I_{3} &
    \\
      & & & 8
    \end{pmatrix}.
\label{eq:lambdamatrix} 
\end{align} 
From the fact that $M_{f_{0}}V = U^{-1} \lambda$, we can now extract the linearly independent basis. Specifically, the new real-space basis vectors are given by
\begin{align}
M_{\text{AS}}V =  \begin{pmatrix}
    \boldsymbol{a}'_{1} & \boldsymbol{a}'_{2} & \boldsymbol{a}'_{3} & \boldsymbol{a}'_{4} & \boldsymbol{a}'_{5} & \boldsymbol{a}'_{6} & \boldsymbol{a}'_{7} & \boldsymbol{a}'_{8}
    \end{pmatrix},
\label{eq:Mf0new} 
\end{align} 
where $\{\boldsymbol{a}'_{i}\}$ are column vectors rotated by the transformation matrix $V$ from $\{\boldsymbol{a}_{i}\}$. The new momentum-space basis vectors are given by 
\begin{align}
U^{-1} = \begin{pmatrix}
    \boldsymbol{b}'_{1} & \boldsymbol{b}'_{2} & \boldsymbol{b}'_{3} & \boldsymbol{b}'_{4} & \boldsymbol{b}'_{5} & \boldsymbol{b}'_{6} & \boldsymbol{b}'_{7} & \boldsymbol{b}'_{8}
    \end{pmatrix},  
\label{eq:Ubasis}  
\end{align}
where $\{\boldsymbol{b}'_{j}\}$ are column vectors rotated by $U^{-1}$ from $\{\boldsymbol{b}_{j}\}$ at $j$=TRIMs. Since the two sets of new bases are related by 
\begin{equation}
\boldsymbol{a}'_{i} = \boldsymbol{b}'_{i} \lambda_{i},
\end{equation}
where $\lambda_{i}$ denotes the diagonal element of $\lambda$, we can span the 0D invariant group for atomic superconductors $ \{ \text{AS} \}$ and $\mathcal{K}$ in the same set of linearly independent bases. 

The explicit form of the symmetry indicators is given by
\begin{equation}
\boldsymbol{\kappa} \equiv U \bar{{\bf{n}}},
\label{eqn:firstnu}
\end{equation}
where $\bar{{\bf{n}}}$ is the set of 0d invariants. There is however a basis ambiguity in calculating the Smith normal form:
\begin{eqnarray}
\lambda &=& U M_{\text{AS}} V 
\\
&=& (U L^{-1}) (L M_{\text{AS}} R^{-1}) (RV),
\label{eqn:LR}
\\
&=& \tilde{U} \tilde{M}_{\text{AS}} \tilde{V}.
\end{eqnarray}
While the symmetry indicator group remains unchanged, the explicit form of the symmetry indicators is now given by
\begin{equation}
\tilde{\boldsymbol{\kappa}} \equiv \tilde{U} \bar{{\bf{n}}}.
\label{eqn:indicator_new}
\end{equation}
Therefore, there is no unique explicit expression for the symmetry indicators without further input. To fix a canonical basis, we need to match with the real-space classification, and we choose a basis such that the strong, mixed, and weak phases are all separated.

\subsection{Ab-initio calculations}
\label{sec:DFT}
Our DFT calculation is performed for centrosymmetric bulk $\beta$-MoTe$_2$ (nonsymmorphic space group \#11 $P2_1/m$,
point group C$_{\rm{2h}}$) with the experimental geometry \cite{Dawson1987}, in the absence of spin-orbit coupling, using VASP 
\cite{VASP1,VASP2}. The experimental lattice constants and angles are as follows \cite{Dawson1987}:
$a$=6.330, $b$=3.469, $c$=13.860 angstrom, $\beta$=93.55$^{\circ}$, and $\alpha$=$\gamma$=90$^{\circ}$. The real space lattice vectors are
$\vec{a}_1=a\hat{e}_x$,~$\vec{a}_2=b\hat{e}_y$,~and $\vec{a}_3=c \: {\rm{cos}}\beta \hat{e}_x + c \: {\rm{sin}}\beta \hat{e}_z$, where $\hat{e}_{x,y,z}$
are unit vectors in Cartesian coordinates. There are four Mo atoms and eight Te atoms in a monoclinic primitive unit cell. The inversion
center of our atomic coordinates in the primitive unit cell is set to the origin. We use projector-augmented wave (PAW) pseudopotentials
\cite{Blochl1994} within the Perdew-Burke-Ernzerhof (PBE) generalized gradient approximation \cite{Perdew1996}. Each Mo atom has 6
valence electrons which are nominally singly occupied at the five $4d$ orbitals and at the $5s$ orbital. Each Te atom has 6 valence
electrons which are nominally occupied at the $5p$ orbitals and $5s$ orbitals. The Hubbard $U$ value of 3.0~eV is used for the Mo $4d$ orbitals, 
within the DFT+U method \cite{Liechtenstein1995} as implemented in VASP, following Ref.~\onlinecite{Exp_MoTe2}. We sample $k$ points of $10\times20\times5$ with $\Gamma$ point centered for the self-consistent calculation. The cutoff of the kinetic energy is set to 400 eV. We consider 54 bands.

In order to compute the topological indices $Z_8$, $Z_4$, and $Z_2$ \cite{SJHuang2021} or the indices in Eq.~(\ref{eq:indicators}), we calculate parity values of all bands at the eight TRIM $k$ points, from the wave function of the self-consistent calculation using two different codes \cite{github_parity,ATLee2014}. The TRIM points are $\Gamma=(0.0, 0.0, 0.0)$, $X=(0.5, 0.0, 0.0)$, $U=(0.5, 0.0. 0.5)$,
$Z=(0.0, 0.0, 0.5)$, $Y=(0.0, 0.5, 0.0)$, $S=(0.5, 0.5, 0)$, $T=(0.0, 0.5, 0.5)$, and $R=(0.5, 0.5, 0.5)$. Here the coordinates of
the $k$ points are in terms of the reciprocal lattice vectors $\vec{b}_1=(\hat{e}_x - cot\beta \hat{e}_z)2\pi/a$, $\vec{b}_2=2\pi/b \hat{e}_y$,
and $\vec{b}_3=2\pi/c \hat{e}_z$. Note that the MoTe$_2$ crystal has nonsymmorphic group. The atoms
in the crystal have twofold screw symmetry along the $y$ axis, $t(\vec{b}_2/2)C_{2y}$ (where $t(\vec{b}_2/2)$ is a translation along
the $y$ or ${b_2}$ axis by $\vec{b}_2/2$), mirror plane $t(\vec{b}_2/2)\sigma_{xz}$ about the $xz$ plane, and inversion symmetry.
The twofold screw symmetry gives twofold degeneracy at $k=\pi/b$ plane in the absence of spin-orbit coupling and each degenerate band consists of
a band with positive parity and a band with negative parity \cite{Matsugatani2021}. Therefore, there is always twofold degeneracy at
the $Y$, $R$, $S$, and $T$ points with an equal number of positive-parity bands and negative-parity bands.
For visualization of the Fermi surfaces, we use the {\tt c2x} program \cite{Rutter2017} and the XCrysDen program \cite{Kokalj1999}.

\begin{figure}[htb]
\begin{center}
\includegraphics[width=0.45 \textwidth]{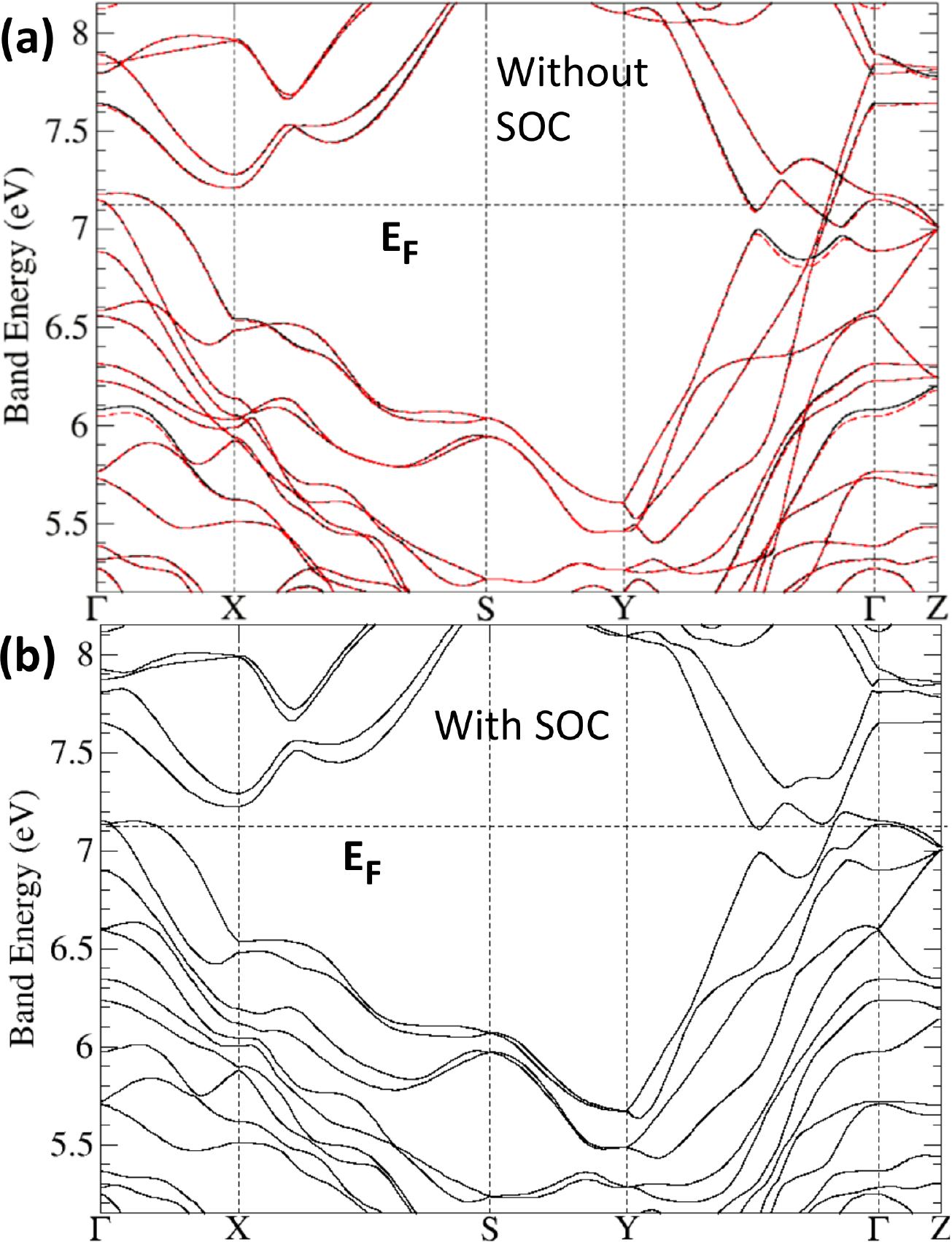}
\caption{(Color online) (a) Band structure obtained using VASP (solid black) and the 44-band tight binding model (dashed red) without spin-orbit coupling where the Fermi level is 7.1515 eV. (b) Band structure using VASP with spin-orbit coupling where the Fermi level is 7.1544 eV. In (a) and (b), the Fermi level is indicated as the horizontal line.}
\label{fig:TB-bands}
\end{center}
\end{figure}

\subsection{Construction of the 44-band tight binding model}
\label{sec:tb}
From the above VASP calculation, we then compute hopping integrals and construct a tight-binding model based on 44 Wannier functions (WFs),
using WANNIER90 code version 1.2 \cite{Marzari1997,Souza2001,Mostofi2008}. The 44 Wannier functions consist of $d_{xy}$, $d_{yz}$, $d_{xz}$, $d_{x^2-y^2}$, and $d_{z^2}$ orbitals
of all four Mo sites and $p_x$, $p_y$, and $p_z$ orbitals of all eight Te sites in the unit cell. We use the same number of $k$ points as
the VASP calculation and set the minimum energy of disentanglement as zero. Only disentanglement \cite{Souza2001} is applied without maximum
localization of the WFs. All disentagled WFs are centered at the atomic sites. We exclude the bottommost eight bands from the VASP calculation
in order to generate the WF-44 tight-binding model. These eight bands have the same numbers of positive parity bands and negative parity bands.
Fewer numbers of WFs than 44 orbitals would produce neither atomic-orbital-shaped WFs nor poor agreement with the VASP band structure. Fig.~\ref{fig:TB-bands}a shows the comparison between the DFT band structure (solid black) and the 44-band tight-binding model (dashed red). They are in good agreement with each other. This band structure is similar to the band structure with spin-orbit coupling (Fig.~\ref{fig:TB-bands}b).

\bibliography{MoTe2.bib}
\end{document}